\documentclass[11pt]{article}
\usepackage[latin1]{inputenc}
\usepackage[english]{babel}
\usepackage[namelimits]{amsmath}
\usepackage{authblk}
\usepackage{amssymb}
\usepackage{amsmath}
\usepackage{amsthm}
\usepackage{graphicx}

{\tiny}

\begin{document}
\title{{\bf } Corrections to the Black hole entropy from a Bose Einstein condensate: a semi-classical  phenomenological approach}
\author[1,2]{S. Viaggiu\thanks{s.viaggiu@unimarconi.it}}
\affil[1]{Dipartimento di Scienze Ingegneristiche, Universit\'a degli Studi Guglielmo Marconi, Via Plinio 44, I-00193 Roma, Italy.}
\affil[2]{Dipartimento di Matematica, Universit\`a di Roma ``Tor Vergata'', Via della Ricerca Scientifica, 1, I-00133 Roma, Italy.}
\date{\today}\maketitle

\begin{abstract} 
In this paper we obtain logarithmic corrections to the black hole entropy. 
Motivated by our recent proposal concerning the nature of 
the degrees of freedom leading to the black hole entropy in terms of a Bose Einstein (BEC) condensate of gravitons, we study how to introduce logarithmic corrections. In fact we show that, after modifying the internal energy by means 
of simple by physically sound arguments dictated by ordinary quantum mechanics and possible
non-commutative effects at Planckian scales, a logarithmic term does appear in 
the Bekenstein Hawking entropy law. We also obtain that the entropy $S_{BH}$ of a ball of Planckian 
areal radius is $2\pi K_B$, i.e. $S_{BH}(R=L_P)=2\pi K_B$.
Our approach show that the possibility that the interior of a black hole is composed with a BEC of gravitons is a viable physically motivated possibility.  
\end{abstract}

{\it Keywords}; Black Holes; logarithmic corrections, Bose-Einstein condensate; gravitons.\\
PACS Number(s): 05.20.-y,04.70.-s,04.60.-m,03.75.Nt

\section{Introduction}

As well known, Hawking discovered \cite{1} that black holes are equipped, thanks to the emission of
a termal radiation, with a non vanishing entropy \cite{1},\cite{2} 
$S_{BH}=K_B A_h/L_P^2$, with $K_B$ the Boltzmann constant, $L_P$ the Planck length and $A_h$ the area of the event horizon. Despite this important and beatiful result, the statistical origin of the degrees of freedom leading to the formula $S_{BH}=K_B A_h/L_P^2$ is still an open problem
(see for exazmple \cite{3,4,5,6,7,8,9,10,11,12,13,14,15,16,17,18}). Motivated by this line 
of research, in \cite{19,20,21,22} we have proposed a statistical description of the black hole entropy in terms of trapped gravitons representing a radiation field with a discrete spectrum.
In the aforementioned papers, in particular in \cite{19,20}, we shown that the one-quarter area law can be obtained within an error of approximatively $10\%$. Recentely, in \cite{23} we have obtained 
the exact one-quarter area law in terms of a BEC of trapped gravitons at the temperature $T_i=0$.
The results in \cite{23} are in agreements with the finding in \cite{24,25,26,27}.
See also \cite{27Q} for the use of a BEC of a boson gas for Schwarzschild and Kerr 
black holes.\\ 
The black hole entropy is semi-classical and thus has been obtained by adopting quantum arguments in the classical background of general relativity. Moreover, the corrections to the semi-classical expression of black hole entropy are an important ingredient that should be incorporated in any 
viable quantum gravity theory.\\
As well known \cite{28}, for any given system with positive specific heat $C$ with 
entropy $S_0$, thermal fluctuations lead to a corrected entropy $S_c$ given by
$S_c=S_0+1/2\ln(S_0)$. However, for a black hole $C<0$ and quantum fluctuations are expected.
It is expected that thermal fluctuations produce positive logarithmic corrections to entropy, while quantum fluctuations generate negative logarithmic corrections \cite{29}. Hence, for an 
asymptotically flat black hole logarithmic corrections are expected such as 
$S_{BH}=S_{0BH}-a_0\ln(S_{0BH})$ with $S_{0BH}=K_B A_h/L_P^2$. The constant $a_0$ is model
dependent and is $a_0=3/2$ in string theory \cite{11,31} and $a_0=1/2$ in loop quantum
gravity context \cite{32}. As a result, it is thus generally believed 
\cite{29,11,14,31,32,33,4,36,37} 
that quantum fluctuations generate corrections to the black hole entropy given by
\begin{equation}
S_{BH}=K_B\frac{A}{4L_P^2}-K_B a_0\ln\left(\frac{A}{4L_P^2}\right)+const.+Higher\;\;\;orders
,\;\;a_0\in\Re.
\label{1}
\end{equation}
Higher order corrections on the thermodynamics of charged ADS black holes and rotating 
and charged BTZ black holes can be found repectively in \cite{R1} and \cite{R2}. Moreover, note that the parameter $a_0$ in (\ref{1}) has been introduced as a free parameter 
for the first time in \cite{R3}.\\
The aim of this paper is to obtain formula (\ref{1}) starting from the results present in 
\cite{23}.\\
In section 2 we present some result present in \cite{23}. In section 3 we show a possible way to introduce logarithmic corrections, while in section 4 we calculate our expression for black hole entropy together with corrections. Section 5 is devoted to the explanation of the role of 
the $\alpha$ parameter in our model. Finally, section 6 is devoted to some conclusions and final remarks.

\section{Preliminaries: the interior of a black hole with a kind of BEC condensate of gravitons} 
 
In \cite{23} we have outlined the black hole entropy by considering a gas of trapped gravitons inside 
the event horizon. As stressed in \cite{23}, there our aim is not obviously to present a naive way to quantize radiation inside a black hole. Conversely, our purpose is to obtain a physically motivated expression for a possible gravitational radiation inside a spherical box (mimicking the interior of a black hole) and thus calculate the resulting entropy with the constraints dictated by general relativity for a black hole.\\
To start with, we consider massless excitations (gravitons) with equation of state $PV=U/3$, where $P$ is the internal pressure of the radiation field, $V=4\pi/3 R^3$ is the so called 
thermodynamic volume \cite{38}, where $R=2GM/c^2$ is the Schwarzschild radius of the black hole,
i.e. of our confining box
and $U=Mc^2=\frac{c^4 R}{2G}$, with $M$ corresponding to the ADM mass of the black hole.
For the internal temperature $T_i$ of the confining box we adopt the expression 
$T_i=\alpha T_{BH}$, where $\alpha\in\Re$ and 
where $T_{BH}=\frac{c\hbar}{4\pi K_B R}$ is the Hawking temperature that is the temperature of the termal radiation emitted from the black hole. From the first law:
\begin{equation} 
T_i dS_{BH}=dU+PdV.
\label{1a}
\end{equation}
With the aforementioned setups, we deduce from (\ref{1}) that the general linear equation of state
\begin{equation} 
PV=\gamma U,\;\;\gamma\in\Re,
\label{1b}
\end{equation}
can be satisfied with $T_i=\alpha T_{BH}$, where $\alpha-1=3\gamma$. Hence a radiation field 
$\gamma=1/3$ can be obtained only with $\alpha=2$.
The further step is to consider a physically motivated expression for the angular frequency 
$\omega$ of trapped gravitons inside a spherical box. This can be obtained by applying 
Dirichlet boundary conditions such that \cite{19,23}
$Z^{(a)}_{\ell m}(R, \omega)=0$, where $Z^{(a)}_{\ell m}(R, \omega)$ is the Regge-Wheeler function
\cite{39} for a traveling gravitational wave in the vacuum (for more details
see \cite{19,23}). We thus have:
\begin{equation}
{\omega}_{n \ell}\simeq\frac{c}{2R}\left(2+\ell+2n\right)\pi,\,\;\ell\geq 2,\;\;n\in\mathbb{N}.
\label{2}
\end{equation}
The partition function for a radiation field of $N$ excitations is given by 
$Z_T=Z_r^N$ with:
\begin{equation}
Z_r=\frac{e^{-\left(\frac{2\pi c\beta\hbar}{R}\right)}}
{\left[1-e^{-\left(\frac{\pi c\beta\hbar}{2R}\right)}\right]
\left[1-e^{-\left(\frac{\pi c\beta\hbar}{R}\right)}\right]},
\label{3}
\end{equation}
where $\beta=1/(K_B T)$. For
$U_r$, after using $-N{\left(\ln Z_r\right)}_{,\beta}$ and subtracting the zero point energy
we obtain
\begin{equation}
U_r=
\frac{c\pi N\hbar}{2R\left[e^{\left(\frac{\pi c\beta\hbar}{2R}\right)}-1\right]}+
\frac{c\pi N\hbar}{R\left[e^{\left(\frac{\pi c\beta\hbar}{R}\right)}-1\right]}.
\label{4}
\end{equation}
As usual entropy is defined as
\begin{equation}
S=-K_B\sum_{\ell=2}^{\infty}\sum_{n=0}^{\infty} P_{n\ell}\ln\left(P_{n\ell}\right),\;\;
P_{n\ell}=\frac{e^{-\beta E_{n\ell}}}{Z_T},
\label{5}
\end{equation}
where $E_{n\ell}=\hbar{\omega}_{n\ell}$ and with ${\omega}_{n\ell}$ given by (\ref{2}).\\
Moreover we have that $TS=NK_B T\ln\left(Z_r\right)+U$, and with 
$F=U-TS$, we obtain $F=-K_B T\ln\left(Z_T\right)$ and hence 
$\frac{\partial F}{\partial T}=-S$ and $\frac{\partial F}{\partial V}=-P$. Consequently for the 
entropy $S$ we get:
\begin{equation}
S=-NK_B\left[\ln\left(1-e^{-\frac{X}{2}}\right)+
\ln\left(1-e^{-X}\right)\right]+
\frac{c\pi\hbar N}{2TR}\frac{\left(3+e^{\frac{X}{2}}\right)}{\left(e^X-1\right)},\;\;
X=\frac{c\pi\beta\hbar}{R}.
\label{6}
\end{equation}
It is easy to see (see \cite{19,23}) that with \cite{2,3,4} we have $PV=U_r/3$
($\gamma=1/3$ or $\alpha=2$), i.e. the equation of state suitable for a radiation field. After imposing the constraint 
$U_r=Mc^2=\frac{c^4 R}{2G}$ dictated by black hole physics, we have:
\begin{equation}
N=\frac{R^2}{\pi L_P^2}
\frac{1}{\frac{1}{\left[e^{\frac{2\pi^2}{\alpha}}-1\right]}+
\frac{2}{\left[e^{\frac{4\pi^2}{\alpha}}-1\right]}}.
\label{7}
\end{equation}
Finally, after putting (\ref{7}) in (\ref{6}) for the black hole entropy $S_{BH}$ 
we get
\begin{eqnarray}
& & S_{BH}=K_B Y(\alpha)\frac{A_h}{4L_P^2},\;\;A_h=4\pi R^2,\nonumber\\
& & Y(\alpha)=\frac{b}{\alpha\pi^2\left(3+e^{\frac{2\pi^2}{\alpha}}\right)}\label{8}\\
& & b=-\alpha e^{\frac{4\pi^2}{\alpha}}\ln\left(1-e^{-\frac{2\pi^2}{\alpha}}\right)-
\alpha e^{\frac{4\pi^2}{\alpha}}\ln\left(1-e^{-\frac{4\pi^2}{\alpha}}\right)+6\pi^2+\nonumber\\
& & +2\pi^2 e^{\frac{2\pi^2}{\alpha}}+\alpha \ln\left(1-e^{-\frac{2\pi^2}{\alpha}}\right)+
\alpha\ln\left(1-e^{-\frac{4\pi^2}{\alpha}}\right)\nonumber. 
\end{eqnarray}
Moreover, we have $Y(\alpha)=\frac{2}{\alpha}+positive\;\;\; terms$ and in order to obtain 
the $1/4$ factor in the entropy we must have $Y(\alpha)=1$. Since we have a radiation field for 
$\alpha=2$ ($\gamma=1/3$), from (\ref{8}) it is follows that for such a value we obtain 
$Y(2)=1+positive\;\;\; terms$. In order to solve this issue, in \cite{23} we considered a different equation of state. We must thus generalize the angular spectrum (\ref{2}) suitable for a 
radiation field. To this purpose,
we introduce a scalar field $\Phi(R)$ with
${\omega}={\omega}_{r}+\frac{\Phi(R)}{N}$.
Consider a radiation field made of massless excitations with angular frequency ${\omega}_{r}$ and energy $E=\hbar{\omega}_r$. The following theorem 
\cite{21,22,23} holds:\\

{\bf Theorem 1}: {\it Let $\{\omega_r\}$ denote the spectrum of a radiation field with equation of state $PV=U_r/3$ composed of $N$
massless excitations enclosed in a spherical box of proper areal radius $R$ and volume $V$.
Then the ones with angular frequency ${\omega}={\omega}_{r}+\frac{\Phi(R)}{N}$ and given internal energy 
$U(R)=Mc^2=U_r+\hbar\Phi(R)$ 
have a linear equation of state $PV=\gamma U,\;\gamma\in\Re$ if and only if the differentiable funcion $\Phi(R)$ satisfies the following equation}:
\begin{equation}
\hbar\left[R\frac{d\Phi}{dR}+\Phi(R)\right]=U(R)(1-3\gamma),
\label{13} 
\end{equation}
{\it together with the condition}
\begin{equation}
U-\hbar\Phi(R)\geq 0.
\label{14}
\end{equation}
For a proof see \cite{21,22}. As shown in \cite{23}, by performing the same calculations leading to
(\ref{8}) and adopting Theorem 1,  with the expression for $S_{BH}$ given by 
(\ref{6}) still holding, for the entropy $S_{BH}$ we have
\begin{equation}
S_{BH}=K_B\frac{A_h}{4L_P^2}\left(1+\frac{\alpha}{2}+o(\alpha)\right),\;\;
\alpha\rightarrow 0.
\label{15}
\end{equation}
with $Y(\alpha)$ given by (\ref{8}). Hence the expression 
$S_{BH}=K_B\frac{A_h}{4L_P^2}$ can be obtained if and only if $\alpha\rightarrow 0$, i.e.
$T_i\rightarrow 0$.\\
In the next sections we show as to obtain logarithmic corrections to black hole entropy by 
performing semi-classical arguments and with a small modification of Theorem 1. 

\section{Generating quantum corrections for $U$}

In order to obtain expression (\ref{15}), an important constraint imposed by general relativity is provided by $U(R)=Mc^2$ that denotes that the energy of the black hole is provided by its ADM energy, i.e. the classical expression for the total mass-energy of the black hole. In order to obtain corrections to black hole entropy in our semi-classical phenomenological approach we must modify
expression $U(R)=Mc^2=\frac{c^4 R}{2G}$. In the following we justify the expression for the modifird intrnal energy in three different ways: with arguments related to possible non-commutative quantum gravity effects, with dimensional arguments and finally with an analogy with the Casimir effects.\\
Tostart with, in \cite{20} we attempted to build corretions to black hole 
entropy in ultraviolet Planckian limit in terms of a possible non-commutativity at the Planckian length
\cite{40} (see also \cite{41,42,43} for a generalization of the uncertainty relations (STUR) present in \cite{40}). In \cite{21,22} we attempted to build corrections to the entropy by adopting reasonings that are consequence of the STUR in \cite{40,41,42} and as a result we obtain that spheres  filled with a matter field with $\gamma\leq -1/3$ (soft matter) are possible only for microscopic Planckian spheres (see \cite{21,22}). Results in \cite{20,21,22} are based on the fact that 
corrections are due to possible non-commutative effects that are expected to be dominant very near the singularity where $r\sim\L_P$, with the singularity substituted by a Planckian ball with Planckian density. Moreover, Planckian limitations in 
\cite{21,22} for spheres filled with $\gamma\leq -1/3$ are based on the fact
that constant present in $\Phi(R)$ is independent on $\gamma$. In the following we show that in our case the constant present in $\Phi(R)$ must be dependent on $\alpha$ and consequently also macroscopic configurations are allowed.\\ 
To start with, we consider similar arguments present in \cite{41}. The purpose is to obtain a suitable physically motivated expression for the energy $E=U$ corrected 
by quantum gravity motivated effects, with $E=U=Mc^2+E_c$, where $M$ is again the ADM mass and 
$E_c$ depicts quantum corrections. 
In order to obtain a suitable expression dor $E_c$, we consider the following  "gedanken experimenten" (see \cite{41}): we must localize a given spacetime spherical region $S$ of radial areal radius $\Delta R$. To localize the aforementioned region, we must concentrate 
an amount $\Delta E$ of energy within $S$. This can be accomplished provided that no black hole 
is formed during the experiment with the usual condition:
\begin{equation}
\frac{G\Delta E}{c^4}\leq\frac{\Delta R}{2}.
\label{16}
\end{equation}
The well known time-energy uncertainty relation states that $\Delta t\Delta E\geq \frac{\hbar}{2}$, that in turns together with inequality (\ref{16}) it gives:
\begin{equation}
c\Delta t\Delta R\geq L_P^2.
\label{17}
\end{equation}
Inequality (\ref{17}) suggests that we can promote $R$ and $t$ to quantum operators and consequently
the (\ref{17}) denotes spacetime uncertainty relation for a non commutative spacetime restricted to spherical localizing states. More precisely, time and radial coordinates become self-adjoint operators acting on some Hilbert space $H$. Hence the short notation $\Delta O$ denotes the uncertainty of the operator $O$ in some quantum state $\omega$ ($\Delta O={\Delta}_{\omega}O)$.
In this regard, a vector state $\omega(O)$ is defined as 
$\omega(O)=\left(\phi, O\phi\right),\;\phi\in H$, where $(.,.)$ indicates scalar product
and consequently, as usual, 
$\Delta O={\Delta}_{\omega}(O)=\sqrt{\omega(A^2)-\omega{(A)}^2}$.
In \cite{40} a complete theory is developed for a non-commutative spacetime in a flat background, with uncertainty relations generalized in \cite{41} and in \cite{42,43} for curved spacetimes.
Moreover, in \cite{40} a state of optimal localization ${\omega}^*$ is a (Gaussian) state such that STUR are exactly satured. For (\ref{17}) we have
\begin{equation}
c{\Delta}_{\omega^*}t{\Delta}_{\omega^*} R=L_P^2. 
\label{18}
\end{equation}
Moreover, for a localizing state with spherical symmetry, as in \cite{40}, we must have coordinates uncertainties of the same magnitudo, i.e. 
$c{\Delta}_{\omega^*} t\sim {\Delta}_{\omega^*} R$ and from (\ref{18}) we obtain
${\Delta}_{\omega^*} R= L_P$ (minimum uncertainty for spherical localizing states). 
Moreover, from ${\Delta}_{\omega^*} t{\Delta}_{\omega^*} E=\frac{\hbar}{2}$ with 
$c{\Delta}_{\omega^*} t\sim {\Delta}_{\omega^*} R$ we thus have:
\begin{equation}
{\Delta}_{\omega^*} E\sim c\frac{\hbar}{2{\Delta}_{\omega^*}R}.
\label{19}
\end{equation}
From (\ref{19}) note that if we consider a sphere of Planckian size, we have that energy fluctuations are maximum at Planck length and are of the order of Planck energy 
$E_P=\sqrt{c^5\hbar/G}$. Hence, expression (\ref{19}) motivates,
according to \cite{21,22}, with ${\Delta}_{\omega^*} E\sim E$ and 
${\Delta}_{\omega^*}R\sim R$,
the following expression for
$E=U=Mc^2+E_c$:
\begin{equation}
U=\frac{c^4 R}{2G}+\frac{Kc^4}{2GR}L_P^2,
\label{20}
\end{equation}
Note that for $R\sim L_P$ the two terms in (\ref{20}) have the same magnitudo thus denoting the fact that quantum corrections are expected to be important at Planckian scales and negligible at larger scales.\\
Another possibility to justify the (\ref{20}) is the following. 
On general ground, we can suppose to have for the ADM mass-energy modified by quntum effects (fluctuations) the expression $U=U_c(R,\hbar)$. We thus also expect that
$\lim_{\hbar\rightarrow 0} U_c=\frac{c^4 R}{2G}$ and that for $R>>L_P$ we have
$U_c\sim \frac{c^4 R}{2G}$ (boundary condition). We can thus consider the following Taylor expansion of $U_c$ with respect to $\hbar$:
\begin{equation}
U_c=\frac{c^4 R}{2G}+\sum_{i=1}^{\infty}{\hbar}^i H_i(R),
\label{RR1}
\end{equation}
together with $\lim_{R\rightarrow\infty} H_i=0$. At the first order in $\hbar$ we have:
\begin{equation}
U_c=\frac{c^4 R}{2G}+\hbar H_1(R)+o(\hbar).
\label{RR2}
\end{equation}
From a dimensional study and thanks to the boundary condition at spatial infinity for $H_i$ we 
can adopt the following expression for $H_i(R)$:
\begin{equation}
H_i(R)=K_i\frac{G^{i-1}}{c^{3i-4}}R^{(1-2i)}.
\label{RR3}
\end{equation} 
After setting $i=1$ from (\ref{RR3}) we obtain $H_1=K_1\frac{c\hbar}{R}$ that is in agreement with 
the correction in (\ref{20}) with $\frac{K}{2}=K_1$. Higher order corrections can be outlined for 
$i>1$. As an example, with $i=2$ we have a further correction 
\begin{equation}
U=\frac{c^4 R}{2G}+\frac{Kc^4}{2GR}L_P^2+\frac{K_2 c^4}{2G R^3}L_P^4+o(\hbar^2),
\label{RR4}
\end{equation}
leading to a sub-leading correction\footnote{Similar corrections can be found in 
\cite{R4}} with respect to the first correction 
to the entropy scaling as $1/\sqrt{A_h}$ (by using Theorem 1).\\
A further way to justify formula (\ref{20}) is to use a possible analogy with Casimir effect
\cite{R5}. As well known, Casimir effect implies the existence of an attractive force between two
perfectly conducting uncharged parallel plates of area $A$ at a given distance $d$. The most widely accepted physical explanation of the Casimir effect is in terms of quantum vacuum corrections representing the
zero point energy of a quantized field. For the energy $E_c$ we have the following formula:
\begin{equation}
E_c=-\frac{c\pi^2\hbar A}{720 d^3}.
\label{RR5}
\end{equation}
In our case we have a spherical symmetry with a positive energy. Hence we are temped to write down
the (\ref{RR5}) with the plus sign and $A\sim d^2$. Fortunately, as well know, the Casimir effect depends on the geometry of the physical configuration \cite{R6}. In fact, for a spherical 
uncharged sphere the force becomes repulsive and $E_c$ positive with $E_c\sim 1/d$, where
$d$ is the radius of the conducting sphere. This fact is in perfect agreement with the
resonings related to non-commutative effects that are expected to produce quantum fluctuations 
with a repulsive force.\\  
Concerning the positive 
constant $K$, on physical ground we expect that $K$ depends on the ratio 
$T_i/T_{BH}$, i.e. $\alpha$ in such a way that $\lim_{\alpha\rightarrow 0}K=0$ and 
as a result
the lowest energy state is formally provided, in the BEC state, by the ADM mass $M$. 
Neverthless, in the next section it will be shown that in that limit the term proportional 
to $K$ in (\ref{20}) does contribute to the corrected entropy provided that 
$K\sim\alpha$. In this regard, we pose $K=\alpha D, D>0$. For more details see the dicussion in section 5.
The positive constant $D$ will be estimated in the next section.\\ 
As explained in the 
previous section, a radiation field with $\gamma=1/3$ ($\alpha=2$) is not suitable in order to obtain 
the one-quarter entropy area law \cite{23}. To this purpose, we can use a small modification of Theorem 1. We start again with a radiation field of massless excitations with angular frequency ${\omega}_{r}$ and energy $E=\hbar{\omega}_r$. We have:\\

{\bf Theorem 2}: {\it Let $\{\omega_r\}$ denote the spectrum of a radiation field with equation of state $PV=U_r/3$ composed of $N$
massless excitations enclosed in a spherical box of proper areal radius $R$ and volume $V$.
Then the ones with angular frequency ${\omega}={\omega}_{r}+\frac{\Phi(R)}{N}$ and given internal energy 
$U(R)=Mc^2+\frac{\alpha D c^4}{2GR}L_P^2=U_r+\hbar\Phi(R)$ 
have a linear equation of state $PV=\gamma U,\;\gamma\in\Re$ if and only if the differentiable funcion $\Phi(R)$ satisfies the following equation}:
\begin{equation}
\hbar\left[R\frac{d\Phi}{dR}+\Phi(R)\right]=U(R)(1-3\gamma),
\label{21} 
\end{equation}
together with condition (\ref{14}).\\
The proof is left unchanged with respect to the one in 
\cite{21}.  Hence, Theorem 1 continues to hold with $U(R)=Mc^2$ substituted by (\ref{20}), i.e.
$U(R)=Mc^2+\frac{Kc^4}{2GR}L_P^2$ and $U_r$ still provided by (\ref{4}). The next step is to calculate $\Phi(R)$ by means of Theorem 2:
\begin{equation}
\hbar\Phi(R)=
\frac{c^4}{4G}R(1-3\gamma)+\frac{c \alpha D\hbar(1-3\gamma)}{2R}\ln\left(\frac{R}{L}\right).
\label{22} 
\end{equation}
First of all note that for a radiation field with $\gamma=1/3$ we have $\Phi(R)=0$ and treatment of previous section holds.
Also note that $L$ is an integration constant. In \cite{21,22}, without loss of generality, it has been chosen  
of the order of Planck length with $L=sL_P, s\sim 1$. The constant $s$ plays no role in the following calculations: thanks to the equation ${\Delta}_{\omega^*} R= L_P$ we can 
set $s=1$.\\ 
We are now ready to calculate the corrected black hole entropy.

\section{The final expression of the corrected black hole entropy}

To start with, thanks to (\ref{22}), for $U_r=U(R)-\hbar\Phi(R)$ we get:
\begin{equation}
U_r=\frac{\alpha c^4 R}{4G}+\frac{\alpha D c^4}{2GR}L_P^2-
\frac{c \alpha D\hbar(1-3\gamma)}{2R}\ln\left(\frac{R}{L_P}\right).
\label{23}
\end{equation} 
Condition (\ref{14})($U_r\geq 0$), after setting $W=R/L_P$, becomes:
\begin{equation}
\alpha\left(\frac{W}{2}+\frac{D}{W}-\frac{D}{W}(2-\alpha)\ln\left( W\right)\right)\geq 0.
\label{24}
\end{equation}
Thanks to the reasonings of section above, we are interested in the limit for $\alpha\rightarrow 0^+$, 
and thus $\alpha>0$ in our parametric region of interest. It is simple to verify that, with $D$ of order of unit or less, the (\ref{24}) is satisfied for $\alpha\geq 0$ and consequently also 
macroscopic configurations ($R>>L_P$) can de depicted. In the treatment present in \cite{21,22}
the constant $K$ in (\ref{20}) is independent on $\alpha$ and consequently from (\ref{24})
we have $R\leq \sqrt{e}L_P$. Also note that if $K$ is chosen independent on $\alpha$, then from 
(\ref{23}) we have that $\lim_{\alpha\rightarrow 0} U_r=\frac{K c^4}{2GR}L_P^2-
\frac{cK\hbar}{R}\ln\left(\frac{R}{L_P}\right)$ and as a result we cannot obtain
the leading term $K_B A_h/L_P^2$ for black hole entropy and in this limit the entropy is diverging.
This further motivates both physically and mathematically the choice 
$K=\alpha D$.\\
From (\ref{23}) with (\ref{4}) we obtain for $N$, instead of (\ref{7}), 
the following equation:
\begin{eqnarray}
& & N=
\frac{B}{\frac{1}{\left[e^{\frac{2\pi^2}{\alpha}}-1\right]}+
\frac{2}{\left[e^{\frac{4\pi^2}{\alpha}}-1\right]}}, \label{25}\\
& & B=\frac{A_h}{4L_P^2}\left(\frac{\alpha}{2\pi^2}\right)+
\frac{\alpha D}{\pi}\left[1+\left(1-\frac{\alpha}{2}\right)\ln(\pi)\right]-
\frac{\alpha D}{\pi}\left(1-\frac{\alpha}{2}\right)
\ln\left(\frac{A_h}{4L_P^2}\right). \nonumber
\end{eqnarray}
Concerning the expression (\ref{6}) for $S_{BH}$, it remains left unchanged also after the introduction of $\Phi(R)$: in terms of $\alpha$ we have 
\begin{equation}
S=-NK_B\left[\ln\left(1-e^{-\frac{2\pi^2}{\alpha}}\right)+
\ln\left(1-e^{-\frac{4\pi^2}{\alpha}}\right)\right]+
NK_B\frac{2\pi^2}{\alpha}e^{-\frac{2\pi^2}{\alpha}},
\label{26}
\end{equation}
where we have used the relation $T_i=\alpha T_{BH}$. 
The expression of $\Phi(R)$ enters crucially in the expression (\ref{25}) for $N$.
The final step is to substitute the expression of $N$ given by (\ref{25}) and (\ref{26}) and 
performing the limit for $\alpha=0$. The final expression for the corrected entropy in 
the limit $\alpha=0$ is:
\begin{equation}
S_{BH}=K_B\left[\frac{A_h}{4L_P^2}-2\pi D\ln\left(\frac{A_h}{4L_P^2}\right)
+2\pi D\left(1+\ln(\pi)\right)\right].
\label{27}
\end{equation}
The entropy (\ref{27}) is nothing else but the leading term of the usual form (\ref{1}).
With a simple physical argument we can also fix the constant $D$. In fact, we expect, as 
stated below equation (\ref{20}), that quantum corrections are maximum at Planckian scales and in particular, as stated below equation (\ref{18}) with ${\Delta}_{\omega^*} R= L_P$. Notice that logarithmic 
corrections have a minus sign in (\ref{27}) ($D>0$), and thus we have that at $R=L_P$ expression 
(\ref{27}) must have an absolute minimum. After setting $x=\frac{A_h}{4L_P^2}$, we have such a minimum, after deriving
the (\ref{27}) with respect to $x$, at $x=2\pi D$. Since the minimum for $R$ is 
$R=L_P$, we obtain $D=\frac{1}{2}$. The (\ref{27}) becomes:
\begin{equation}
S_{BH}=K_B\left[\frac{A_h}{4L_P^2}-\pi\ln\left(\frac{A_h}{4L_P^2}\right)
+\pi\left(1+\ln(\pi)\right)\right].
\label{28}
\end{equation}
Notice that $S_{BH}(R=L_P)=2\pi K_B$, i.e. twice the value without corrections. It is 
interesting to compare the value $a_0=\pi$ that we have obtained (see equation (\ref{1})) 
with the ones obtained in string ($a_0=3/2$) and loop ($a_0=1/2$) context. To this purpose, 
we suppose that the minimum for $R$ is attained at $R_{m}=sL_P$. With the same reasonings above we 
get that $a_0=3/2$ is obtained with $s=\sqrt{3/(2\pi)}$, while $a_0=1/2$ is attained 
with $s=\sqrt{1/(2\pi)}$: in both cases, $R_m<L_P$. These reasonings can indicate that the 
value $a_0=\pi$ we have obtained can be a feature due to a possible non-commutative origin of 
logartithmic corrections  of black hole entropy. 

\section{The role of the parameter $\alpha$}

In section 3 we have setted $K=\alpha D$, where $T_i=\alpha T_{BH}$ and $\alpha-1=3\gamma$.
We have also shown that for $K$ independent on $\alpha$, the entropy expression
(\ref{6}) is diverging in the limit for $\alpha\rightarrow 0$ that in turn is an essential 
request in order to obtain the leading term $S_{BH}=K_B A_h/L_P^2$ (see expression (\ref{15})).
Apart from this mathematical necessity, one may ask what is the physical meaning of the limit 
$\alpha\rightarrow 0$.\\
To start with, consider expression (\ref{20}) with $K=\alpha D$. In the limit for $D=0$
we regain expression (\ref{15}), while with the further limit $\alpha=0$ we obtain the usual expression
$S_{BH}=K_B A_h/L_P^2$ for entropy.
However, note that, after setting $D=0$, apart from the terms depending on $R$,
in the limit for $\alpha=0$ the dominant term of $N$ leading to (\ref{15}) is (see 
\cite{23} for more details):
\begin{equation}
N\sim \alpha e^{\frac{2\pi^2}{\alpha}}.
\label{A1}
\end{equation}
Hence we have that, at fixed $R$, $\lim_{\alpha\rightarrow 0} N=+\infty$.\\ 
This is reminescent of thermodynamic
limit where statistical fluctuations are zero in the limit for
$N\rightarrow\infty$.\\ 
There, the important point is that $N$ is diverging for $\alpha=0$. Moreover,
note that the function $\alpha e^{\frac{2\pi^2}{\alpha}}$ in 
(\ref{A1}),
has an absolute minimum for $\alpha=2\pi^2$ and is monotonically decreasing for
$\alpha< 2\pi^2$ and it is thus formally invertible with respect to $\alpha$ leading
to the expresion $\alpha\sim F(N,R)$ with $\lim_{N\rightarrow\infty} F(N,R)=0$.
The parameter $\alpha$ drives the behavior of statistical fluctuations.
To a better understanding of this fact, consider the same calculation leading 
to the (\ref{15}) but with $D\neq 0$ from the onset. After repeating the same caclulations of the case $D=0$, instead of the (\ref{15}) we obtain, at the first order with respect to $\alpha$:
\begin{eqnarray}
& & S_{BH}=S_{BH}(\alpha=0)+\frac{\alpha A_h}{8\pi^2 L_P^2}+D\alpha
\left(\frac{\ln\pi}{\pi}+\frac{1}{\pi}-\pi\ln\pi\right)+\nonumber\\
& & +\pi\alpha D\left(1-\frac{1}{\pi^2}\right)\ln\frac{A_h}{4L_P^2}+o(\alpha), \label{A2}
\end{eqnarray}
where $S_{BH}(\alpha=0)$ is nothing else but the (\ref{27}).
Also in this case, we have an expression similar to the (\ref{A1}) with a $R$ dependence including logarithmic corrections. As a consequence the results are left unchanged with 
$\alpha\sim Q(N,R)$ with $\lim_{N\rightarrow\infty} Q(N,R)=0$, where obviously $Q$ is different 
from $F$ with $\lim_{D\rightarrow 0}Q=F$. The statistical nature of fluctuations depicted by 
$\alpha$ is confirmed by an inspection of formula (\ref{A2}). In fact, note that logarithmic correction
proportional to $\alpha$ in (\ref{A2}) is 
\begin{equation}
+\pi\alpha D\left(1-\frac{1}{\pi^2}\right)\ln\frac{A_h}{4L_P^2}
\label{A3}
\end{equation}
and thus the coefficient of such a correction is positive. As mentioned at the introduction,
logarithmic corrections with positive sign \cite{28} are due to thermal fluctuations, while the ones with negative logarithmic corrections are due to quantum fluctuations \cite{29}.\\
Summarizing, the parameter $\alpha$ drives the behavior of thermal fluctuations that in turn are 
vanishing for $\alpha=0$, i.e. $N\rightarrow\infty$ and $T_i\rightarrow 0$ with only 
quantum fluctuations surviving in such a limit. 

\section{Conclusions and final remarks}

In this paper we continued the investigation present in \cite{23} with the purpose to obtain the well known logarithmic corrections of black hole entropy. This task is obtained by proposing a generalization of the ADM mass-energy $M$ of a black hole in terms of fluctuations motivated by
possible Planckian effects. In fact, by inspection of (\ref{28}) or (\ref{1}), it is easy to see
that logarithmic corrections are of the same magnitudo of the leading term
$K_B\frac{A_h}{4L_P^2}$ only at Planckian scales. For an ordinary macroscopic black hole
with $R\sim 10^3$ meters, we have that 
$K_B\frac{A_h}{4L_P^2}\sim 10^{74} K_B$, while 
$K_B \ln\left(\frac{A_h}{4L_P^2}\right)\sim 262K_B$. More generally, the ratio between the two aforementioned terms, after setting $R=\beta L_P$, looks like 
${\beta}^2/(\ln(\pi {\beta}^2))$ and thus grows
very quickly for $\beta>>1$. We also obtain that $S_{BH}(R=L_P)=2\pi K_B$, i.e. the entropy of
a Planckian sphere is not vanishing and represents the smallest sphere we can consider in a 
non-commutative scenario.\\
The generalization of ADM mass-energy has been also justified in terms of dimensional arguments and by an analogy with the Casimir effect.\\
Our approach is in some sense phenomenological since the 
generalization of $U$ including possible quantum effects is motivated by sound arguments of general relativity and quantum mechanics. As a result, we obtain, according to
\cite{23}, a BEC of gravitons at $T_i=0$ but with the logarithmic corrections 
to black hole entropy.\\
As a final consideration we could generalize our treatment to a Kerr black hole with mass $M$ and
angular parameter $a$ by using the same transformations used at the end of paper \cite{20}. This 
could be interesting for example to study a realistic situation with a rotating black hole surrounded
by a galactic halo \cite{44} . This can be matter for further investigations.

\end{document}